# Nanoscopic Characterization of DNA within Hydrophobic Pores: Thermodynamics and Kinetics

Fernando J.A.L. Cruz,[1,2,*] Juan J. de Pablo[2,3] and José P.B. Mota[1]

[1]*LAQV, REQUIMTE, Departamento de Química, Faculdade de Ciências e Tecnologia, Universidade Nova de Lisboa, Caparica, 2829-516, Portugal.E-mail: fj.cruz@fct.unl.pt*
[2]*Department of Chemical and Biological Engineering, University of Wisconsin-Madison, Madison, Wisconsin, 53706, USA*
[3]*Institute of Molecular Engineering, University of Chicago, Chicago, Illinois, 60637, USA*

**The energetic and transport properties of a double-stranded DNA dodecamer encapsulated in hydrophobic carbon nanotubes are probed employing two limiting nanotube diameters, $D = 4$ nm and $D = 3$ nm, corresponding to (51,0) and (40,0) zig-zag topologies, respectively. It is observed that the thermodynamically spontaneous encapsulation in the 4 nm nanopore ($\Delta G \approx -40$ kJ/mol) is annihilated when the solid diameter narrows down to 3 nm, and that the confined DNA *termini* directly contact the hydrophobic walls with no solvent slab in-between. During the initial moments after confinement ($t \leq 2$–3 ns), the biomolecule translocates along the nanopore's inner volume according to Fick's law ($\sim t$) with a self-diffusion coefficient $D = 1.713 \cdot 10^{-9}$ m$^2$/s, after which molecular diffusion assumes a single-file type mechanism ($\sim t^{1/2}$). As expected, diffusion is anisotropic, with the pore main axis as the preferred direction, but an in-depth analysis shows that the instantaneous velocity probabilities are essentially identical along the $x$, $y$ and $z$ directions. The 3D velocity histogram shows a maximum probability located at $<v> = 30.8$ m/s, twice the observed velocity for a single-stranded three nucleotide DNA encapsulated in comparable armchair geometries ($<v> = 16.7$ m/s, $D = 1.36$–1.89 nm). Because precise physiological conditions (310 K and [NaCl] = 134 mM) are employed throughout, the present study establishes a landmark for the development of next generation *in vivo* drug delivery technologies based on carbon nanotubes as encapsulation agents.**

## 1. INTRODUCTION

A plethora of applications currently envisage carbon nanotubes (CNTs) as next-generation encapsulation media for biological polymers, such as proteins and nucleic acids [1, 2]. Owing to several appealing features, such as large surface areas, well-defined physico-chemical properties and the hydrophobic nature of their pristine structure, CNTs are considered ideal candidates to be used as nanopores for biomolecular confinement. Present day potential applications span different purposes and objectives, such as intracellular penetration via endocytosis and delivery of biological cargoes [3-5], ultrafast nucleotide sequencing [6-8] and gene and DNA delivery to cells [5, 9, 10]. The remarkable experimental work by Geng *et al.* [5, 11] has shown that carbon nanotubes can spontaneously penetrate the lipid bilayer of a liposome, and the corresponding hybrid incorporated into live mammalian cells to act as a nanopore through which water, ions and DNA are delivered to the cellular interior. For an efficient and cost-effective industrial fabrication of SWCNT-based technology for DNA encapsulation/delivery, the interactions between the solid and the biomolecule need to be thoroughly understood in order to render the DNA/SWCNT device able to be used under physiological conditions, $T = 310$K and [NaCl] = 134 mM. Nonetheless, the energetics and dynamics of single- (ssDNA) and double-stranded DNA (dsDNA) encapsulation onto single-walled carbon nanotubes (SWCNTs) are virtually unexplored and the corresponding molecular level details remain rather obscure. Previous theoretical and experimental work with DNA and SWCNTs has fundamentally been focused on the solids' external volume, overlooking the possibility of molecular encapsulation [12-14]; nevertheless, it is well known that the conformational properties of biopolymers under confinement are of crucial relevance in living organisms (*e.g.*, DNA packaging in eukaryotic chromosomes, viral capsids). Unrealistic high temperature (400 K) studies revealed that depending on pore diameter, a small 8 nucleobase-long ssDNA strand can be spontaneously confined [15]. However, there is a critical diameter of 1.08 nm [16], bellow which molecular confinement is inhibited by an energy barrier of *ca*. 130 kJ/mol, arising essentially from strong van der Waals repulsions [17]. These findings were extended for intratubular confinement of a 2 nm long ssDNA onto a SWCNT mimic of the constriction region of an α-hemolysin channel [18].

dsDNA confinement in carbon nanotubes remains utterly uncharted, for most of the earlier work has focused on temperatures remarkably distinct from the physiological value, thus preventing extrapolation of results to *in vivo* conditions. The pioneer work of Lau *et al.* [19] showed that a small dsDNA molecule (8 basepairs long), initially confined onto a $D = 4$ nm diameter nanotube, exhibits a root-mean squared displacement similar to the unconfined molecule, but that behaviour is drastically reduced as the nanotube narrows to $D = 3$ nm; Cruz *et al.* [20, 21] have already demonstrated that diffusion inside SWCNTs can exhibit deviations from the classical Fickian behaviour. The insertion of dsDNA onto multi-walled carbon nanotubes has been experimentally observed by STM/STS, TEM and Rahman techniques [22, 23]. However, it seems to be a competing mechanism with the wrapping of the biomolecule around the nanotube external walls; Iijima's reported data failed to identify the relevant conditions upon which the confinement process is favoured, such as ionic strength of the media and temperature [22]. In order to probe the thermodynamical spontaneity of encapsulation, we have adopted the well-known Dickerson dodecamer [24] as dsDNA model (Figure 1) and conducted a series of well-tempered metadynamics calculations [25] involving two distinct nanotubes, namely (51,0) and (40,0) with skeletal diameters of $D = 4$ nm and $D = 3$ nm, respectively; very importantly, the results were obtained under precise physiological conditions, and the media ionic strength maintained at 134 mM by employing a NaCl buffer.

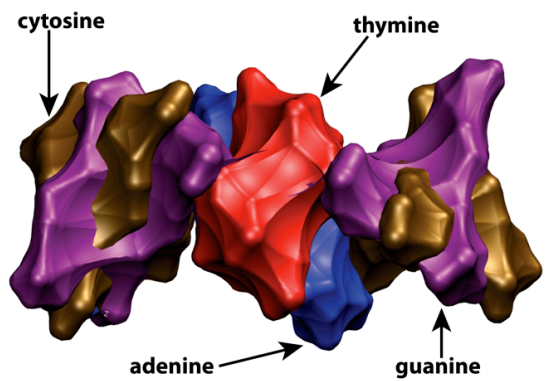

**Figure 1 – Dickerson dsDNA dodecamer.** Isovolumetric representation of the B-DNA Dickerson dodecamer [24] with sequence 5'-D(*CP*GP*CP*GP*AP*AP*TP*TP*CP*GP*CP*G)-3', length $L = 3.8$ nm and skeletal diameter $D = 2$ nm, highlighting the changing chemical nature along the double strand; nucleobase residues are coloured according to their chemical nature, namely blue (A), red (T), purple (G) and brown (C). Note that the whole DNA molecule is atomistically detailed in the calculations, and thus each individual atom has a corresponding partial electrostatic charge.

Our work indeed shows that the dsDNA molecule, initially in a bulk solution, can become encapsulated onto a $D = 4$ nm SWCNT leading to a pronounced decrease of the whole system's Gibbs free-energy [26]. The de Gennes blob theory for polymers has been extended by Jun *et al.* [27] to include the effect of cylindrical confinement upon the biopolymer free energy, and the latter decreases with an increase of nanotube diameter; because the blob description breaks down once the diameter approaches the DNA persistence length (strong confinement), Dai *et al.* [28] recently extended Jun's formalism to dsDNA strongly confined in slit-pore geometries. In order to address some of these encapsulation issues, the present work provides a full thermodynamical mapping of the associated Gibbs surface as well as a structural and kinetic analysis of dsDNA within cylindrical nanopores. The remainder of the manuscript is organized as follows: molecular models and methods are described in Section 2, followed by a discussion of the main results obtained (Section 3) and finally highlighting some conclusions and future lines of work (Section 4).

## 2. METHODOLOGY AND ALGORITHMS

### 2.1 Molecular Models

Molecules are described using atomistically detailed force fields, including electrostatic charges in each atom. The dispersive interactions are calculated with the Lennard-Jones (12,6) potential, cross parameters between unlike particles determined by the classical Lorentz-Berthelot mixing rules, and electrostatic energies described by Coulomb's law. DNA is treated as a completely flexible entity within the framework of the AMBER99sb-ildn force-field [29, 30], the corresponding potential energies associated with bond stretching, $U(r)$, and angle bending, $U(\theta)$, are calculated with harmonic potentials, whilst the dihedral energies are computed using Ryckaert-Bellemans functions,
$$U(\varphi) = \sum_{dihedrals} \sum_{n=0}^{5} C_n [\cos(\varphi - 180°)]$$
. To retain computational tractability, we have chosen the double-stranded B–DNA Dickerson dodecamer [24], exhibiting a pitch [31] of $P \sim 3.4$ nm obtained from an average of 10–10.5 base-pairs per turn over the entire helix [32], and with a double-strand end-to-end length of $L \sim 3.8$ nm measured between terminal (GC) base pairs (Fig.1); the A–DNA form has $P \sim 2.6$ nm corresponding to an average of 11 base-pairs per turn [32]. Considering that the B–DNA backbone P atoms lie on a cylindrical surface, the diameter of the double-strand corresponds to $D \sim 2$ nm [31]. In spite of smaller in length than genomic DNA, the Dickerson dodecamer main structural features resemble those of genomic λ-bacteriophage DNA [33], namely in the radius of gyration and double-strand backbone diameter, $R_g \approx 0.7–1$ nm and $D \approx 2$ nm. The Na$^+$ and Cl$^-$ ions are described with the parameterization of Aqvist and Dang [34] and the H$_2$O molecules by the TIP3P force field of Jorgensen and co-workers [35].

Large diameter ($D \approx 4$ nm) SWCNTs have been recently prepared by Kobayashi *et al.* [36]. In order to examine DNA confinement into such large, hollow nanostructures, two different diameter SWCNTs were adopted, both with zig-zag symmetry and length $L = 8$ nm; skeletal diameters, measured between carbon atoms on opposite sides of the wall, are $D = 4$ nm (51,0) and $D = 3$ nm (40,0). The solid walls are built up of hexagonally-arranged $sp^2$ graphitic carbon atoms, with a C–C bond length [21, 37] of 1.42 Å, whose Lennard-Jones potential is given by Steele's parameterization ($\sigma = 0.34$ nm, $\varepsilon = 28$ K) [38].

### 2.2 Molecular Dynamics and Metadynamics

Molecular dynamics (MD) simulations in the isothermal–isobaric ensemble ($NpT$) were performed using the Gromacs set of routines [39]. Newton's equations of motion were integrated with a time step of 1 fs and using a Nosé-Hoover thermostat [40, 41] and a Parrinello-Rahman barostat [42] to maintain temperature and pressure at 310 K and 1 bar. A potential cut-off of 1.5 nm was employed for both the van der Waals and Coulombic interactions, and the long-range electrostatics were calculated with the particle-mesh Ewald method [43, 44] using cubic interpolation and a maximum Fourier grid spacing of 0.12 nm. Three-dimensional periodic boundary conditions were applied throughout the systems.

The well-tempered metadynamics scheme of Barducci and Parrinello [25] was employed to obtain the free-energy landscape associated with the confinement mechanism. Briefly, the well-tempered algorithm biases Newton dynamics by adding a time-dependent Gaussian potential, $V(\zeta,t)$, to the total (unbiased) Hamiltonian, preventing the system from becoming permanently trapped in local energy minima and thus leading to a more efficient exploration of the phase space. The potential $V(\zeta,t)$ is a function of the so-called collective variables (or order parameters), $\zeta(q) = [\xi_1(q), \xi_2(q), \ldots, \xi_n(q)]$, which in turn are related to the microscopic coordinates of the real system, $q$, according to

$$V(\zeta(q),t) = W \sum_{t'=0}^{t' \leq t} \exp\left(-\frac{V(\zeta(q(t')),t')}{\Delta T}\right) \exp\left(-\sum_{i=1}^{n}\frac{(\xi_i(q)-\xi_i(q(t')))^2}{2\sigma_i^2}\right) \quad (1)$$

where $t$ is the simulation time, $W=\tau_G \omega$ is the height of a single Gaussian, $\tau_G$ is the time interval at which the contribution for the bias potential, $V(\zeta,t)$, is added, $\omega$ is the initial Gaussian height, $\Delta T$ is a parameter with dimensions of temperature, $\sigma_i$ is the Gaussian width and $n$ is the number of collective variables in the system; we have considered $\tau_G = 0.1$ ps, $\omega = 0.1$ kJ/mol, $\Delta T = 310$ K and $\sigma = 0.1$ nm. The parameter $\Delta T$ determines the rate of decay for the height of the added Gaussian potentials and when $\Delta T \to 0$ the well-tempered scheme approaches an unbiased simulation. Because SWCNTs are primarily one-dimensional symmetric, the free-energy landscape was constructed in terms of two collective variables, $\xi_1 = \vec{R}_z^{DNA} - \vec{R}_z^{SWCNT}$ and $\xi_2 = |\vec{R}_1^{GC} - \vec{R}_{12}^{GC}|$, where $\vec{R}$ is the positional vector of the centre of mass of the biomolecule ($\vec{R}_z^{DNA}$) and of the nanotube ($\vec{R}_z^{SWCNT}$), projected along the z–axis, or of the terminal (GC) nucleobase pairs at the double-strand *termini*, ($\vec{R}_1^{GC}$) and ($\vec{R}_{12}^{GC}$). According to our definition of collective variables, $\xi_1$ corresponds to the z-distance between the biomolecule and the nanopore centre and $\xi_2$ can be interpreted as the DNA end-to-end length. The characteristic lengths of the nanotube and Dickerson dodecamer are, respectively, $L = 8$ nm and $L = 3.8$ nm, and therefore any value $\xi_1 = \Delta L = (L^{SWCNT} - L^{DNA})/2 < 2.1 nm$ corresponds to a DNA–SWCNT hybrid in which the biomolecule is completely encapsulated within the solid boundaries; the threshold $\xi_1 > 5.9$ nm obviously indicates the absence of confinement. At the end of the simulation, the three-dimensional free-energy surface is constructed by summing the accumulated time-dependent Gaussian potentials according to $F(\zeta,t) = -\frac{T+\Delta T}{\Delta T}V(\zeta,t)$. A discussion of the algorithm's convergence towards the correct free-energy profile is beyond the scope of this work and can be found elsewhere [25, 45]; suffices to say that it in the long time limit $\partial V(\zeta,t)/\partial t \to 0$ and therefore the well-tempered method leads to a converged free-energy surface. An alternative approach to obtain the time-independent free-energy surface relies on integrating $F(\zeta,t)$ at the final portion of the metadynamics run. The converged free-energy can thus be mathematically obtained from equation (2):

$$F(\zeta) = -\frac{1}{\tau}V \int_{t_{tot}-\tau}^{t_{tot}} V(\zeta,t)dt \quad (2)$$

where $t_{tot}$ is the total simulation time and $\tau$ is the time window over which averaging is performed. We have implemented a convergence analysis for each collective variable, $\xi_1$ and $\xi_2$, by splitting the last 40 ns of simulation time into $\tau = 4$ ns windows, and confirming the convergence of the bias potential $V(\zeta, t)$. This test ensures that the Gibbs map of Figure 2 is a good estimator of the free-energy changes associated with molecular encapsulation.

### 2.3 Statistical Fittings of Velocity profiles

Velocity data were histogram reweighted with a bin width of 0.002 nm/ps (Figure 6) and analysed in terms of the Gram-Charlier peak function (eqn. 3) or a sigmoidal logistic function type 1 (eqn. 4), where $p(v)$ is the probability and $v$ is the velocity measured along the corresponding direction.

$$p(v) = p_0 + \frac{A}{w\sqrt{2\pi}} e^{-z^2/2}\left(1 + \left|\sum_{i=3}^{4}\frac{a_i}{i!}H_i(z)\right|\right) \quad v(x,y,z)$$
$$z = \frac{v-v_c}{w}, H_3 = z^3 - 3z, H_4 = z^4 - 6z^3 + 3 \quad (3)$$

$$p(v) = \frac{a}{1 + e^{-k(v-v_c)}} \quad v(x), v(y), v(z) \quad (4)$$

### 3. RESULTS AND DISCUSSION

We have adopted two macroscopic descriptors to construct the free-energy landscape, $FE(\xi_1)$ and $FE(\xi_2)$, relating the centres of mass of DNA and SWCNT projected along the nanopore main axis ($\xi_1$) and the DNA end-to-end length ($\xi_2$) (*cf.* Section 2.2). An inspection of the corresponding Gibbs map (Figure 2) clearly indicates the existence of five distinct free-energy minima, all located sequentially along the nanopore internal volume, $\xi_1 < 1.8$ nm. Because they represent thermodynamically identical free-energy basins, DNA can translate freely within the nanopore, easily moving from one free-energy minimum to the next, while exploring a maximum probability configuration path between adjacent minima. However, in order to irreversibly escape from those deep free-energy valleys, $FE(\xi_1, \xi_2) \sim -40$ kJ/mol$^{-1}$, DNA has to overcome large energetical barriers, rendering the exit process towards the bulk thermodynamically expensive. Furthermore, an inspection of Figure 2 reveals a position where the DNA double-strand located at $\xi_1 = 0.149$ nm corresponds to the absolute free-energy minimum, highlighting the nanopore centre as the thermodynamically favoured region for the encapsulated molecule. The residual free-energy region located between $-0.1 < \xi_1$ (nm) $< 0$ is physically meaningless, for the system is always positioned in a phase space comprised between $\xi_1^{min} = 5.4 \cdot 10^{-6}$ nm and $\xi_1^{max} = 2.9$ nm; the probability distributions associated with $\xi_1$ and $\xi_2$, obtained by histogram reweighting with a minute binwidth of 0.001 nm are recorded in Supplementary Information (Fig. SI1). A detailed convergence analysis of the well-tempered metadynamics technique is beyond the scope of the present work and has been discussed elsewhere [45]. Suffices to say that in the present case the individual free-energy profiles recorded in Figure 2, obtained by 4 ns integration time windows of the free-energy with respect to $\xi_1$, clearly indicate that the technique has converged to the correct Gibbs map; although not shown, a similar conclusion upholds regarding $\xi_2$.

**Figure 2.** *Top)* **Gibbs free-energy landscape of dsDNA@(51,0) SWCNT.** The Gibbs surface is built in terms of two distinct macroscopic descriptors, $\xi_1$ and $\xi_2$; $\xi_1$ is the distance between centres of mass of the dsDNA and nanotube projected along the nanopore's main axis, $\xi_1 = \vec{R}_z^{DNA} - \vec{R}_z^{SWCNT}$, and $\xi_2$ is the absolute dsDNA end-to-end length, $\xi_2 = \left|\vec{R}_1^{GC} - \vec{R}_{12}^{GC}\right|$, where $\vec{R}$ is the positional vector of the centre of mass of the biomolecule ($\vec{R}_z^{DNA}$) and of the nanotube ($\vec{R}_z^{SWCNT}$) or of the terminal (GC) nucleobase pairs at the double-strand *termini*, ($\vec{R}_1^{GC}$) and ($\vec{R}_{12}^{GC}$). Note that the consecutive free-energy minima along $\xi_1$ indicate that dsDNA is free to translate along the nanotube. The snapshots were taken at different time intervals corresponding to ($\xi_1$, $\xi_2$) nm: *A)* (0.149, 4.112), *B)* (0.621, 4.164), *C)* (1.306, 4.164) and *D)* (1.794, 4.115); H$_2$O molecules and Na$^+$ and Cl$^-$ ions have been omitted for clarity sake. *Bottom)* $\xi_1$ **convergence profiles.** Each set of data was obtained by a 4 ns integration time of free-energy regarding $\xi_1$,

$$FE(\xi_1) = -\frac{1}{4}\int_{\tau_j}^{\tau_j+4ns} V(\xi_1,t)dt$$

, performed over the last 40 ns of a simulation run of 70 ns length. Note that the overall $FE(\xi_1)$ profile depicted in black, obtained in $\Delta t = 40-70$ ns, exactly overlaps with the final 4 ns time window (66–70 ns, dark grey). It is clear that the free-energy profiles converged to the corresponding minima after *ca.* 60 ns, with only minor contributions being added from this time onwards; the Gibbs landscape recorded in the top figure is an accurate representation of the thermodynamical free-energy changes associated with the encapsulation process.

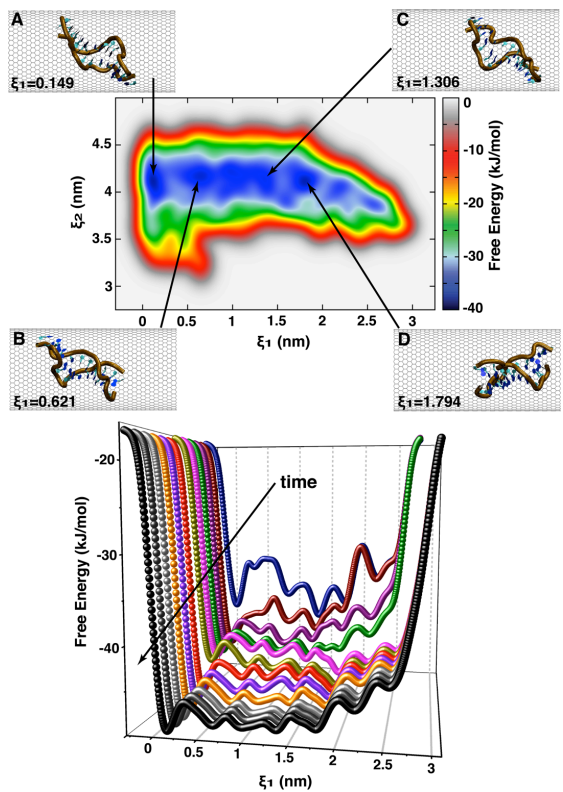

The narrowness of the employed (40,0) topology, $D = 3$ nm, by contrast with the (51,0) SWCNT, seems to prevent confinement even over an observation time window as large as 0.1 µs. Instead of encapsulation, the DNA molecule's direct contact with the (40,0) nanopore leads to the occurrence of two intrinsically distinct situations: *i)* owing to the strong π–π stacking interactions of terminal (GC) pairs with the graphitic mesh, the molecule exoadsorbs and threads the external SWCNT surface according to a mechanism previously observed by Zhao and Johnson [12], or *ii)* the biomolecule becomes trapped at the nanopore entrance, and partial melting of the double-strand *terminii* occurs. Using a (20,20) SWCNT with $D = 2.67$ nm to probe the encapsulation of siRNA, Mogurampelly and Maiti [46] showed that, due to a large free-energy barrier located at the nanopore entrance, confinement was thermodynamically prohibited.

In order to better understand the energetics of interaction between the encapsulated molecule and the solid walls, an analysis of the van der Waals interactions has been implemented and the corresponding findings graphically recorded in Figure 3. Because the DNA centre-of-mass corresponds to the central inner tract formed by four nucleobase pairs (AATT, Fig.1), it is clear that its' interaction with the graphitic walls is rather minimal compared to the energies obtained from the two outer (CGCG) tracts; regardless of the observation time window, the *termini* energies always correspond to more that 87% of the dispersive energy of interaction between the whole DNA molecule and the solid walls, $U^{LJ}$, and, on average correspond to 97%. It is also very interesting to observe that the distance between centres of mass of both molecules is essentially due to translocation of the DNA molecule along the nanopore main axis, for the corresponding 3D distance is either symmetrical about 0 (pore centre) or coincident with its $z$ component (Figure 3); this comes to show that, in spite of the radial symmetry characteristic of nanotubes, the observed molecular mobility within the (51,0) encapsulating volume arises essentially from translation along the nanopore main axis and is severely restricted along the radial direction. Also, and apart from an initial time window of 20 – 22 ns, when the DNA molecule is accommodating itself to the confining volume, and when both energy and distance are considered at the same time, the constancy of the former with the latter demonstrates that the position of the DNA c.o.m. along the main axis does not determines the energy of interaction between biomolecule and solid, e.g., either located at the nanopore centre ($d = 0$ nm) or close to the *termini* ($d = 2 – 2.5$ nm) the van der Waals energy is approximately constant.

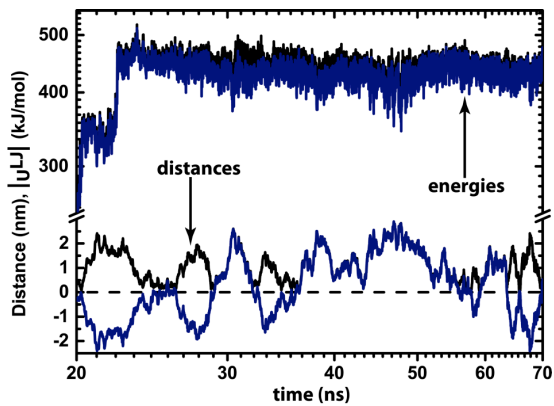

**Figure 3. Dispersive energies and distances between dsDNA and (51,0) SWCNT.** Energies are plotted in terms of their absolute values, $\left| U^{LJ} = U(r) + U(\theta) + U(\varphi) \right|$ for the total interaction between the biomolecule and the solid (black) and the energy term corresponding to the two (CGCG) tracts located at the double strand *termini* (blue). Distances correspond to the 3D distance, $d(x,y,z) = \sqrt{d(x)^2 + d(y)^2 + d(z)^2}$, between c.o.m. (black) and its $z$-component, $d(z)$ (blue). The DNA double-strand has a sequence of 5'-D(*CP*GP*CP*GP*AP*AP*TP*TP*CP*GP*CP*G)-3', and thus the total van der Waals energy plotted in black corresponds to the sum of two terms: obtained from the two (CGCG) tracts, plotted in blue, and from the central (AATT) tract, not shown.

To accurately characterize the structure and dynamics of confined dsDNA within the hydrophobic (51,0) nanopore, independent calculations have been conducted with the metadynamics biasing potential turned off, rendering the simulations exactly equivalent to Newton's law. The local structure associated with the biomolecule was retrieved by calculating the radial distribution function, $g(r) = (V/V_r N_i N_j) \left\langle \sum_i \sum_{j \neq i} \delta(r - r_{ij}) \right\rangle$, where $V$ is the volume of the system, $V_r$ is the volume of a spherical shell at distance $r$ from each particle $i$, $N_i$ is the number of particles $i$ in the system, $r_{ij}$ is the distance between particles $i$ and $j$ and the triangular brackets denote an ensemble average over the entire simulation time window [47, 48]. This procedure was performed for several different $r_{ij}$, namely the dsDNA and the SWCNT centres of mass (c.o.m.), the DNA end-to-end length, and the distance between the terminal (GC) basepairs and the graphitic inner surface. The results recorded in Figure 4 indicate that, upon encapsulation, the DNA's most probable site for occupancy is the nanopore centre, as indicated by the sharp peak located at 0.38 nm in Figure 4A, in striking agreement with the free energy minimum of $\xi_1 = 0.149$ nm (Figure 2). As previously mentioned, the biomolecule is relatively free to travel between adjacent free-energy minima along the $\xi_1$ direction, being able to explore the complete nanopore inner volume as indicated by the 0.4 nm and 3.62 nm grey peaks of Figure 4B: bearing in mind that the (51,0) nanotube employed in the calculations has a length of $L = 8$ nm, the $r = 0.4$ nm signal is equivalent to the pore centre and the $r = 3.62$ nm peak corresponds to one of the two symmetrical pore *termini*, where $L/2 = 4$ nm. The DNA's end-to-end length radial distribution function, measured between opposite (GC) pairs in the double strand, and recorded as black bars in Figure 4B, is well described by a Gaussian statistics with an average peak maximum located at $r = 3.92$ nm highlighting the stability of DNA's B-form characteristic of the Dickerson dodecamer (Figure 1). Furthermore, owing to the strong π–π interactions between the bare (GC) *termini* and the $sp^2$ graphitic surface [37], the former are in direct contact with the latter, corroborating the absence of any solvent (hydrophilic) slab between the graphitic walls and the encapsulated DNA, as indicated by the 0.35 nm signal in Figure 3C; it should be noted that the van der Waals diameter of a graphitic carbon atom corresponds to 0.34 nm. The (40,0) topology, with an effective diameter of $D^{eff} = (3 - 0.34) = 2.66$ nm, becomes too narrow for the DNA molecule to coil and allow its hydrophobic moieties (nucleobases) to directly contact the walls, such as the case for the (51,0) solid ($D^{eff} = 3.66$ nm). Therefore, confinement is annihilated in the (40,0) nanotube because it would lead to an hybrid whose hydrophilic skeleton (phosphates) would be in direct contact with the hydrophobic carbons on the walls.

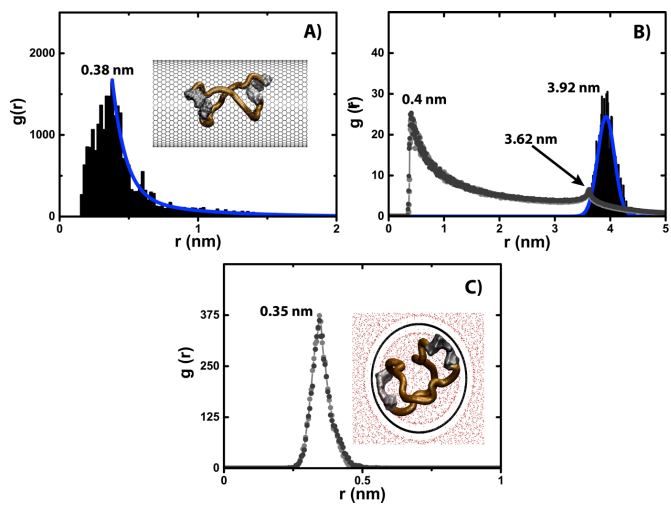

**Figure 4. Radial distribution functions at the (51,0) SWCNT.** *A)* between DNA and SWCNT c.o.m., *B)* between (r$^1$r$^{24}$) and SWCNT c.o.m. (light grey) and (r$^{12}$r$^{13}$) and SWCNT c.o.m. (dark grey) and between (r$^1$r$^{24}$) and (r$^{12}$r$^{13}$) c.o.m. (black), equivalent to the DNA end-to-end length, *C)* between (r$^1$r$^{24}$) and SWCNT surface (light grey) and (r$^{12}$r$^{13}$) and SWCNT surface (dark grey). The blue lines correspond to the best statistical fittings of data with

$$g(r) = A_1 \exp(-r/r_1) + A_2 \exp(-r/r_2) + g^0(r)$$

($A_1$ = 479.44, $r_1$ = 0.53 nm, $A_2$ = 68589.62, $r_2$ = 0.10 nm, $g^0(r)$ = –1.80) (Figure A) and

$$g(r) = \phi \exp\left[-1/2\left(r - r^0/\sigma\right)^2\right]$$

($\phi$ = 24.40, $r^0$ = 3.92 nm, $\sigma$ = 0.15nm) (Figure B). Snapshots were taken to illustrate the local physical environment: *ochre)* DNA individual strands, *grey)* sp$^2$ graphitic mesh and *red)* H$_2$O oxygen atoms.

For any nanotube-based drug delivery technology to find its way into the industrial production line, not only does the encapsulation mechanism needs to be thermodynamically favourable, therefore reducing energy costs, but it also needs to be reversible once the confined genetic material is ready to be delivered to the host cell. The recent theoretical work by Xue *et al.* [49] has demonstrated the feasibility of such ejection process, using filler agents (C60) and mechanical actuators (Ag nanowires) to eject ssDNA from within purely hydrophobic SWCNTs. Between encapsulation and ejection, the nucleic acid needs to travel from entrance to exit, corresponding to opposite nanopore entrances, in order to become available for cellular delivery: what happens in between? That is, how do we characterize the DNA kinetics once it is confined?

One of the most relevant transport properties for industrial applications is the self-diffusion coefficient, *D*, which provides a measure of how mobile a fluid can be once it becomes encapsulated. This property can be determined from a time dependent analysis of molecular trajectories, using either the Stokes-Einstein or the Green-Kubo equations, which have become the *de facto* method [48, 50]; both formalisms are mathematically equivalent [51]. Recently, Baidakov *et al.* [52] determined the self-diffusion coefficient of stable and meta-stable Lennard-Jones fluids via the particles mean-squared displacement (Einstein equation) and velocity auto-correlation functions (Green-Kubo equation), and concluded that both approaches led to self-consistent results within the calculation error (0.5–1.0%). A similar conclusion was obtained by Takeuchi and Okazaki [53] for the self-diffusivities of polymethylene with O$_2$ as a solute.

Transport data were analysed herein by calculating the fluid mean-squared displacement, $MSD = [r(t) - r(0)]^2$, and relating it with the self-diffusion coefficient using the Stokes-Einstein equation, $D = \lim_{t \to \infty} \frac{1}{6t} \langle MSD \rangle$, where $\mathbf{r}(t)$ is the positional vector of a unique molecule at time *t*, and the triangular brackets denote an ensemble average. Special care was taken to accurately sample the MD data; in order to achieve statistically significant results, we used small time delays between origins separated by 1 ps. A further analysis was conducted by decomposing the *MSD* obtained for the DNA center of mass into its *x*, *y* and *z* components. The results recorded in Figure 5A indicate that the biomolecule maintains its translational mobility whilst confined within the SWCNT, visiting a region of space whose boundaries are delimited by the nanopore *termini*, and associated with the adjacent free-energy minima identified in Figure 2 ($\xi_1$ < 1.8 nm). It is very interesting to observe that during an initial time-window of *ca.* 2–3 ns the molecule diffuses according to Fick's law, *e.g.*, the c.o.m. mean-squared displacement is linearly proportional to time, which corresponds to an effective self-diffusion coefficient of $D = 1.789 \cdot 10^{-9}$ m$^2$/s. After that initial time-window, DNA approaches a single-file type diffusion [20], when the *MSD* becomes proportional to $t^{1/2}$; translation within the nanotube is anisotropic. Indeed, the relative occurrences plotted in Figure 5B strongly indicate that molecular displacement along the *z* direction (nanotube main axis) is favoured over the *x* and *y* analogues, corresponding to more than 33 % of the overall (3D) mean-squared displacement. This is clearly due to entropic reasons, for we have shown that the DNA molecule maintains direct contact with the solid walls (Fig. 4C) rendering the *z* direction as the preferred degree of freedom.

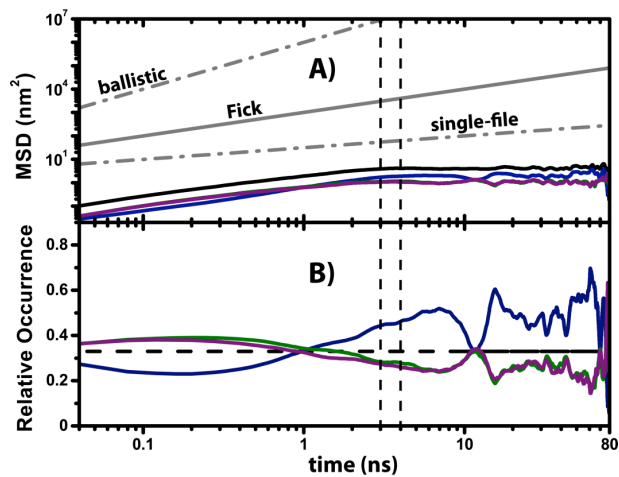

**Figure 5. Kinetics of DNA confined at the (51,0) SWCNT.** *A)* Mean-squared displacement profiles, *MSD*, for the molecular center of mass, and *B)* relative *MSD* occurrence regarding dimension *i*, $MSD^i/MSD^{3D}$. The grey lines correspond to the classical diffusion mechanisms, $MSD \propto t^2$ (ballistic), $MSD \propto t$ (Fick) and $MSD \propto \sqrt{t}$ (single-file) and the dashed black lines are the boundaries of the 3 – 4 ns time window after which molecular diffusion changes from Fickian to single-file type. *black)* Three-dimensional $MSD^{3D}$, *blue)* MSD *z*-component, *green)* MSD *x*-component and *purple)* MSD *y*-component. Notice the anisotropic diffusion of DNA along the nanotube main axis, *z*, responsible for more than 33% of the effective 3D diffusion coefficient.

In order to produce a nanoscopic picture of the DNA's mobility mechanism, the 3D instant c.o.m. velocity was calculated and the corresponding components recorded in Figure 6 after histogram reweighting with a bin width of 0.002 nm/ps. It now becomes clear that, although molecular mobility is anisotropic, favouring DNA translocation along the nanopore main axis, it is mostly due to entropic reasons, because the kinetic energies (velocities) are identical independently of the particular direction in space. The *in vitro* experiments of Geng *et al.* [5] showed that porin-embedded nanotubes are able to transport ssDNA at an average speed of 1.5 nucleotides per millisecond towards the cellular internal volume; on the other hand, Pei and Gao [16] studied the translocation of a small three oligonucleotide ssDNA through armchair geometries, *D* = 1.36 nm and *D* = 1.89 nm, and observed an average translocation velocity of <*v*> = 16.66 m/s. Considering that the overall 3D velocity observed for the dsDNA c.o.m. is reasonably described by Gaussian statistics (Figure 6), a distribution maximum is observed at <*v*>$^{max}$ = 30.8 m/s.

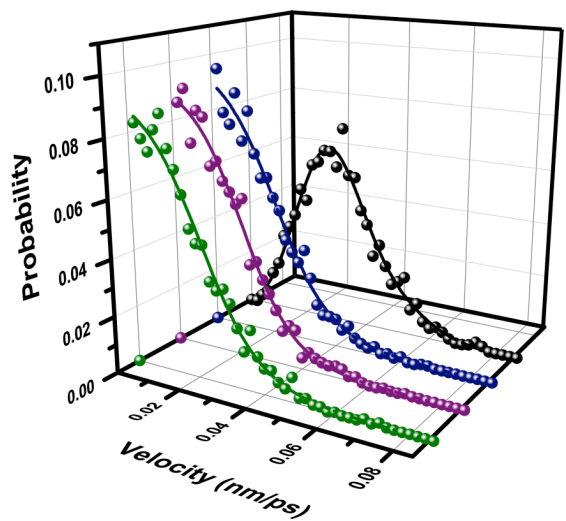

**Figure 6. Velocity probability profiles of confined dsDNA.** Instant velocities were calculated for the entire observation time window (80 ns) and histogram reweighted with a bin width of 0.002 nm/ps: *black)* molecular center of mass (3D), *blue)* *z*-component, *green)* *x*-component and *purple)* *y*-component. Symbols correspond to histogram data and lines are the best statistical fittings (*cf.* Section 2.3); analysis along any individual direction (*x*, *y*, *z*) includes only the decay region.

## 4. CONCLUSIONS AND OUTLOOK

The encapsulation of an atomically detailed DNA dodecamer onto pristine (51,0) carbon nanotubes, *D* = 4 nm, is thermodynamically favourable (Fig.2), resulting in DNA@SWCNT hybrids with lower Gibbs free-energy than the unconfined molecule (ΔGibbs ~ – 40 kJ/mol). Nonetheless, when the nanotube diameter narrows down to a (40,0) topology, *D* = 3 nm, encapsulation is completely inhibited. In this case, the biomolecule either gets stuck at the nanopore entrance or threads the solid outer walls in contact with the solid, as a direct consequence of strong π – π stacking between a terminal (GC) nucleobase pair and the graphitic mesh. Very interestingly, the dispersive energies between DNA and the SWCNT are rather independent of the former's position within the confining volume (Fig.3), and essentially result from the (CGCG) tracts located at the *termini*, *e.g.*, more than 87%. The confined DNA maintains its translational mobility within the pore, <*v*>$^{max}$ = 30.8 m/s, however, due to entropic reasons, translocation is highly anisotropic and results essentially from molecular displacement along the nanotube main axis, as verified by the decomposition between centres of mass distance (Fig.3) and mean-squared displacement (Fig.5) into their corresponding *z* components. This molecular translation occurs between adjacent free-energy minima, all located within the pore inner volume, and corresponding to a DNA conformation similar to the canonical B form.

An inspection of the Gibbs free-energy map obtained for the (51,0) topology indicates a distance between consecutive free-energy minima of 0.47–0.69 nm. This has been overlooked before and seems to suggest the existence of a characteristical length, which, however, we think is unrelated to the graphitic mesh geometry; keeping in the mind the *sp*$^2$ bond between Carbon atoms (0.142 nm), the hexagons composing the solid lattice are roughly 0.38 nm wide. To address this issue, further calculations would be compelling, employing much longer nanotubes and DNA strands. In order to keep the calculations computationally tractable and atomistically detailed, a smaller system than the one being reported now could be employed, either by reducing the simulation box size or decreasing the outer walls solvation slab to focus on the endohedral volume.

#### ACKNOWLEDGEMENTS
This work makes use of results produced with the support of the Portuguese National Grid Initiative (https://wiki.ncg.ingrid.pt). F.J.A.L. Cruz gratefully acknowledges financial support from FCT/MCTES (Portugal) through grant EXCL/QEQ-PRS/0308/2012.

#### REFERENCES

1. Kumar, H.; Lansac, Y.; Glaser, M. A.; Maiti, P. K., Biopolymers in Nanopores: Challenges and Opportunities. *Soft Matter* 2011, 7, 5898–5907.
2. Vashist, S. K.; Zheng, D.; Pastorin, G.; Al-Rubeaan, K.; Luong, J. H. T.; Sheu, F.-S., Delivery of Drugs and Biomolecules using Carbon Nanotubes. *Carbon* 2011, 49, 4077–4097.
3. Canton, I.; Battaglia, G., Endocytosis at the Nanoscale. *Chem. Soc. Rev.* 2012, 41, 2718–2739.

4. Kostarelos, K.; Bianco, A.; Prato, M., Promises, Facts and Challenges for Carbon Nanotubes in Imaging and Therapeutics. *Nature Nano.* 2009, 4, 627-633.
5. Geng, J.; Kim, K.; Zhang, J.; Escalada, A.; Tunuguntla, R.; Comolli, L. R.; Allen, F. I.; Shnyrova, A. V.; Cho, K. R.; Munoz, D.; Wang, Y. M.; Grigoropoulos, C. P.; Ajo-Franklin, C. M.; Frolov, V. A.; Noy, A., Stochastic transport through carbon nanotubes in lipid bilayers and live cell membranes. *Nature* 2014, 514, 612–615.
6. Venkatesan, B. M.; Bashir, R., Nanopore Sensors for Nucleic Acid Analysis. *Nature Nano.* 2011, 6, 615-624.
7. Liu, H.; He, J.; Tang, J.; Liu, H.; Pang, P.; Cao, D.; Krstic, P.; Joseph, S.; Lindsay, S.; Nuckolls, C., Translocation of Single-Stranded DNA Through Single-Walled Carbon Nanotubes *Science* 2010, 327, 64-67.
8. Meng, S.; Maragakis, P.; Papaloukas, C.; Kaxiras, E., DNA Nucleoside Interaction and Identification with Carbon Nanotubes. *Nano Letters* 2007, 7, 45-50.
9. Wu, Y.; Phillips, J. A.; Liu, H.; Yang, R.; Tan, W., Carbon Nanotubes Protect DNA Strands During Cellular Delivery. *ACS Nano* 2008, 2, 2023–2028.
10. Yum, K.; Wang, N.; Yu, M.-F., Nanoneedle: A multifunctional tool for biological studies in living cells. *Nanoscale* 2010, 2, 363–372.
11. Kim, K.; Geng, J.; Tunuguntla, R.; Comolli, L. R.; Grigoropoulos, C. P.; Ajo-Franklin, C. M.; Noy, A., Osmotically-Driven Transport in Carbon Nanotube Porins. *Nano Lett.* 2014, 14, 7051-7056.
12. Zhao, X.; Johnson, J. K., Simulation of Adsorption of DNA on Carbon Nanotubes. *J. Am. Chem. Soc.* 2007, 129, 10438-10445.
13. Alegret, N.; Santos, E.; Rodriguez-Fortea, A.; Rius, F. X.; Poblet, J. M., Disruption of small double stranded DNA molecules on carbon nanotubes: A molecular dynamics study. *Chem. Phys. Lett.* 2012, 525-26, 120-124.
14. Santosh, M.; Panigrahi, S.; Bhattacharyya, D.; Sood, A. K., Unzipping and Binding of Small Interfering RNA with Single Walled Carbon Nanotube: A platform for Small Interfering RNA Delivery. *J. Chem. Phys.* 2012, 136, 65106.
15. Gao, H.; Kong, Y.; Cui, D., Spontaneous Insertion of DNA Oligonucleotides into Carbon Nanotubes. *Nano Letters* 2003, 3, 471-473.
16. Pei, Q. X.; Lim, C. G.; Cheng, Y.; Gao, H., Molecular Dynamics Study on DNA Oligonucleotide Translocation Through Carbon Nanotubes. *J. Chem. Phys.* 2008, 129, 125101.
17. Lim, M. C. G.; Zhong, Z. W., Effects of Fluid Flow on the Oligonucleotide Folding in Single-walled Carbon Nanotubes. *Phys. Rev. E* 2009, 80, 041915.
18. Zimmerli, U.; Koumoutsakos, P., Simulations of Electrophoretic RNA Transport Through Transmembrane
Carbon Nanotubes. *Biophys. J.* 2008, 94, 2546–2557.
19. Lau, E. Y.; Lightstone, F. C.; Colvin, M. E., Dynamics of DNA Encapsulated in a Hydrophobic Nanotube. *Chem. Phys. Letters* 2005, 412, 82–87.
20. Cruz, F. J. A. L.; Müller, E. A., Behavior of Ethylene/Ethane Binary Mixtures within Single-Walled Carbon Nanotubes. 2- Dynamical Properties. *Adsorption* 2009, 15, 13-22.
21. Cruz, F. J. A. L.; Müller, E. A.; Mota, J. P. B., The Role of the Intermolecular Potential on the Dynamics of Ethylene Confined in Cylindrical Nanopores. *RSC Advances* 2011, 1, 270-281.
22. Iijima, M.; Watabe, T.; Ishii, S.; Koshio, A.; Yamaguchi, T.; Bandow, S.; Iijima, S.; Suzuki, K.; Maruyama, Y., Fabrication and STM-characterization of Novel Hybrid Materials of DNA/carbon nanotube. *Chem. Phys. Lett.* 2005, 414, 520.
23. Ghosh, S.; Dutta, S.; Gomes, E.; Carroll, D.; Ralph D'Agostino, J.; John Olson; Guthold, M.; Gmeiner, W. H., Increased Heating Efficiency and Selective Thermal Ablation of Malignant Tissue with DNA-Encased Multiwalled Carbon Nanotubes. *ACS Nano* 2009, 3, 2667–2673.
24. Drew, H. R.; Wing, R. M.; Takano, T.; Broka, C.; Tanaka, S.; Itakura, K.; Dickerson, R. E., Structure of a B-DNA Dodecamer: Conformation and Dynamics. *Proc. Nat. Acad. Sci.* 1981, 78, 2179-2183.
25. Barducci, A.; Bussi, G.; Parrinello, M., Well-Tempered Metadynamics: A Smoothly Converging and Tunable Free-Energy Method. *Phys. Rev. Lett.* 2008, 100, 020603.
26. Cruz, F. J. A. L.; de Pablo, J. J.; Mota, J. P. B., Endohedral Confinement of a DNA Dodecamer onto Pristine Carbon Nanotubes and the Stability of the Canonical B Form. *J. Chem. Phys.* 2014, 140, 225103.
27. Jun, S.; Thirumalai, D.; Ha, B.-Y., Compression and Stretching of a Self-Avoiding Chain in Cylindrical Nanopores. *Phys. Rev. Let.* 2008, 101, 138101.
28. Dai, L.; Jones, J. J.; Maarel, J. R. C. v. d.; Doyle, P. S., A systematic study of DNA conformation in slitlike confinement. *Soft Matter* 2012, 8, 2972-2982.
29. Wang, J.; Cieplak, P.; Kollman, P. A., How Well Does a Restrained Electrostatic Potential (RESP) Model Perform in Calculating Conformational Energies of Organic and Biological Molecules? *J. Comput. Chem.* 2000, 21, 1049–1074.
30. Lindorff-Larsen, K.; Piana, S.; Palmo, K.; Maragakis, P.; Klepeis, J. L.; Dror, R. O.; Shaw, D. E., Improved side-chain torsion potentials for the Amber ff99SB protein force field. *Proteins* 2010, 78, 1950-1958.
31. Franklin, R. E.; Gosling, R. G., Molecular Configuration in Sodium Thymonucleate. *Nature (London)* 1953, 171, 740-741.
32. Vargason, J. M.; Henderson, K.; Ho, P. S., A crystallographic map of the transition from B-DNA to A-DNA. *Proc. Nat. Acad. Sci.* 2001, 98, 7265–7270.
33. Wang, Y.; Tree, D. R.; Dorfman, K. D., Simulation of DNA Extension in Nanochannels. *Macromolecules* 2011, 44, 6594–6604.
34. Noy, A.; Soteras, I.; Luque, F. J.; Orozco, M., The Impact of Monovalent Ion Force Field Model in Nucleic Acids Simulations. *Phys. Chem. Chem. Phys.* 2009, 11, 10596–10607.
35. Jorgensen, W. L.; Chandrasekhar, J.; Madura, J. D.; Impey, R. W.; Klein, M. L., Comparison of Simple Potential Functions for Simulating Liquid Water. *J. Chem. Phys.* 1983, 79, 926-935.
36. Kobayashi, K.; Kitaura, R.; Nishimura, F.; Yoshikawa, H.; Awaga, K.; Shinohara, H., Growth of Large-diameter ( 4 nm) Single-wall Carbon Nanotubes in the Nanospace of Mesoporous Material SBA-15. *Carbon* 2011, 49, 5173–5179.
37. Cruz, F. J. A. L.; de Pablo, J. J.; Mota, J. P. B., Free Energy Landscapes of the Encapsulation Mechanism of DNA Nucleobases onto Carbon Nanotubes. *RSC Advances* 2014, 4, 1310-1321.
38. Steele, W. A., Molecular Interactions for Physical Adsorption. *Chem. Rev.* 1993, 93, 2355-2378.
39. Hess, B.; Kutzner, C.; Spoel, D. v. d.; Lindahl, E., GROMACS 4: Algorithms for Highly Efficient, Load-Balanced, and Scalable Molecular Simulation. *J. Chem. Theory Comp.* 2008, 4, 435–447.
40. Nosé, S., A unified formulation of the constant temperature molecular dynamics methods. *J. Chem. Phys.* 1984, 81, 511-519.
41. Hoover, W. G., Canonical Dynamics: Equilibrium Phase-space Distributions. *Phys. Rev. A* 1985, 31, 1695-1697.
42. Parrinello, M.; Rahman, A., Polymorphic Transitions in Single Crystals: A New Molecular Dynamics Method. *J. Appl. Phys.* 1981, 52, 7182-7190.
43. Darden, T.; York, D.; Pedersen, L., Particle Mesh Ewald: An N log(N) Method for Ewald Sums in Large Systems. *J. Chem. Phys.* 1993, 98, 10089-10092.
44. Essmann, U.; Perera, L.; Berkowitz, M. L.; Darden, T.; Lee, H.; Pedersen, L. G., A Smooth Particle Mesh Ewald Potential. *J. Chem. Phys.* 1995, 103, 8577-8592.
45. Laio, A.; Gervasio, F. L., Metadynamics: a Method to Simulate Rare Events and Reconstruct the Free Energy in Biophysics, Chemistry and Material Science. *Rep. Prog. Phys.* 2008, 71, 126601.
46. Mogurampelly, S.; Maiti, P. K., Translocation and Encapsulation of siRNA Inside Carbon Nanotubes. *J. Chem. Phys.* 2013, 138, 034901.
47. Rowlinson, J. S.; Swinton, F. L., *Liquids and Liquid Mixtures*. Butterworths: London, 1982.
48. Haile, J. M., *Molecular Dynamics Simulation: Elementary Methods*. Wiley: New York, 1992.
49. Xue, Q.; Jing, N.; Chu, L.; Ling, C.; Zhang, H., Release of encapsulated molecules from carbon nanotubes using a displacing method: a MD simulation study. *RSC Adv.* 2012, 2, 6913–6920.
50. Allen, M. P.; Tildesley, D. J., *Computer Simulation of Liquids*. Clarendon Press: Oxford, 1990.
51. Zwanzig, R., Time-correlation functions and Transport Coefficients in Statistical Mechanics. *Annu. Rev. Phys. Chem.* 1965, 16, 67-102.
52. Baidakov, V. G.; Kozlova, Z. R., The self-diffusion Coefficient in Metastable States of a Lennard–Jones Fluid. *Chem. Phys. Letters* 2010, 500, 23–27.
53. Takeuchi, H.; Okazaki, K., Molecular Dynamics Simulation of Diffusion of Simple Gas Molecules in a Short Chain Polymer. *J. Chem.Phys.* 1990, 92, 5643-5652.